\documentclass{aastex}
\usepackage{emulateapj5}
\usepackage{graphicx}
\usepackage{times,psfig}

\def\ltsima{$\; \buildrel < \over \sim \;$}
\def\lsim{\lower.5ex\hbox{\ltsima}}
\def\gtsima{$\; \buildrel > \over \sim \;$}
\def\gsim{\lower.5ex\hbox{\gtsima}}

\newcommand{\be}{\begin{equation}}
\newcommand{\en}{\end{equation}}
\newcommand{\ergs}{\rm \ erg \; s^{-1}}

\def\cmdue {\rm \ cm^{-2}}

\def\msole {~M_{\odot}}

\begin{document}

\title{XMM-Newton observations of IGR J00291+5934: signs of a thermal spectral component 
during quiescence}  

\author{Sergio Campana\altaffilmark{1}, 
Luigi Stella\altaffilmark{2}, Gianluca Israel\altaffilmark{2},
Paolo D'Avanzo\altaffilmark{1}}

\altaffiltext{1}{INAF-Osservatorio Astronomico di Brera, Via Bianchi
46, I--23807 Merate (Lc), Italy}
\altaffiltext{2}{INAF-Osservatorio Astronomico di Roma, Via di
Frascati 33, I-00040 Monteporzio Catone (Roma), Italy}

\email{sergio.campana@brera.inaf.it}


\begin{abstract}
We present X--ray observations of the transient accretion-powered millisecond pulsar 
IGR J00291+5934 during quiescence. IGR J00291+5934 is the first source among 
accretion powered millisecond pulsars to show signs of a thermal component in its 
quiescent spectrum. Fitting this component with a neutron star atmosphere or 
a black body model we obtain soft temperatures ($\sim 64$ eV 
and $\sim 110$ eV, respectively). As in other sources of this class a hard 
spectral component is also present, comprising more than $60\%$ of the unabsorbed 
0.5--10 keV flux. Interpreting the soft component as cooling emission from the 
neutron star, we can conclude that the compact object can be spun up to milliscond 
periods by accreting only $\lsim 0.2\msole$.

\end{abstract}


\section{Introduction}

IGR J00291+5934 (IGR J0029 in the following) is one out of eight low mass neutron 
star transient showing coherent pulsations during outbursts (see Wijnands 2005 for a 
review). 
Distinctive properties of these faint transients are weaker outburst peak
luminosities ($\sim (1-5)\times 10^{36}\ergs$) with respect to classical neutron 
star transients (see Campana et al. 1998 for a review), very low mass companions (mass
functions $<2\times 10^{-3}\msole$), short orbital periods ($P_{\rm orb}\lsim 4.3$ hr),
very faint quiescent X--ray luminosities ($\lsim 5\times 10^{31}\ergs$) and absence of
a soft X--ray spectral component in quiescence (quiescent spectrum consistent only of 
a hard power law component).

IGR J0029 was discovered while in outburst on December 2nd 2004 during a routine monitoring 
of the Galactic plane with the INTEGRAL satellite (Eckert et al. 2004). Follow-up RXTE 
observations revealed 598.88 Hz (1.67 ms) coherent pulsations (Markwardt, Swank \& 
Strohmayer 2004), making it the fastest confirmed accretion-powered millisecond pulsar and a 
2.46 hr orbital modulation (Markwardt et al. 2004). 
In quiescence IGR J0029 was detected by Chandra at the beginning of 2005 (and historically by 
ROSAT; Jonker et al. 2005). The source quiescent 0.5--10 keV unabsorbed flux was 
$\sim 8 \times 10^{-14}$ erg cm$^{-2}$ s$^{-1}$ with a variability among different observations 
by a factor of 2 (Jonker et al. 2005). Chandra detected IGR J0029 also in November 2005 
at a level of $\sim 7 \times 10^{-14}$ erg cm$^{-2}$ s$^{-1}$ (Torres et al. 2008). 
Finally a short observation was carried out with Chandra in September 2006 with the source
at a level of $\sim 1\times 10^{13}$ erg cm$^{-2}$ s$^{-1}$ (Jonker, Torres \& Steeghs 2008).
All these Chandra observations collected a relatively low number of photons and a detailed spectral 
analysis could not carried out, being all spectra consistent with a single power-law (photon index 
$2.4\pm0.5$, $90\%$ confidence level).

The non-detection of X--ray bursts makes the estimate of the distance to IGR J0029 difficult.
Shaw et al. (2005) gave an upper limit of 3.3 kpc considering the source position with respect 
to the Galactic Center. Torres et al. (2008) estimated a distance of $2-4$ kpc based on the critical 
X---ray luminosity needed to ionize the accretion disc and produce the observed X--ray light curve 
during outburst. We will assume in the following a fiducial distance of 3 kpc.

In this letter we report on the first study of IGR J0029 in quiescence with XMM-Newton. 
The larger throughput of the XMM-Newton instruments allowed us to study in detail the 
quiescent spectrum. In Sect. 2 we discuss the XMM-Newton data analysis. Sect. 3
contains our discussion and conclusions.

\section{Data analysis}

The XMM-Newton Observatory (Jansen et al. 2001) includes three 1500 cm$^2$
X--ray telescopes each with a European Photon Imaging Camera (EPIC, 0.1--15
keV) at the focus. Two of the EPIC imaging spectrometers use MOS CCDs (Turner
et al. 2001) and one uses pn CCDs (Str\"uder et al. 2001). 

\begin{table}
\caption{Spectral fits.}
\begin{center}
\begin{tabular}{ccccc}
Instrument& Exp. time&Filt. exp. time& Counts & Backg. Sub. Counts\\
          &  (s)     &  (s)          &(cts)   & (cts)             \\
pn        & 26970    & 17305         & 466    &  269      \\
MOS1      & 28535    & 23920         & 142    &  102      \\
MOS2      & 28535    & 23877         & 146    &  107      \\
\end{tabular}
\end{center}
\end{table}

XMM-Newton observed IGR J0029 on July 24 2007 for 27 ks with the thin filter 
on all the EPIC instruments. Data were processed using SAS v.6.6.0. At the end of 
the observation a strong soft proton flares occurred limiting the usable observing time 
to 17 and 24 for the pn and the MOS detectors, respectively (filtering out 
background flares for total 0.2--15 keV rates less than 5 and 20 c s$^{-1}$ for 
the MOS and pn cameras, respectively).
We extracted the pn source spectrum from a circular region with radius $22''$ and the 
background spectrum from a larger region close to the source with radius $34"$ within the 
same CCD. For the MOS cameras we used the same source extraction region and a $84"$ 
background region. 
We obtained 466, 142 and 146 counts for pn, MOS1 and MOS2 cameras in the 0.2--10 keV range.
The background contribution is at a level of $50\%$ and $30\%$ for the pn and MOS cameras, 
respectively.

We considered for timing analysis only the pn light curve in consideration of the higher 
signal to noise ratio.
We subtracted the background light curve and we did not find any evidence 
for variability down to the 500 s, i.e. the shortest timescale that could be investigated 
(the fit with a constant provides a reduced $\chi^2=0.8$ with 43 degrees 
of freedom, dof). This is at variance with what observed in the optical band, where strong flares 
($\sim 1$ mag) were reported on timescale of $\gsim 500$ s (Jonker et al. 2008).

The spectral analysis was carried out by using the data from the three EPIC cameras together. 
Spectra were binned to, at least, 20 counts per spectral bin resulting in 37 bins.
A single power law fit provides a good description of the data (reduced $\chi^2=0.9$, 34 dof) but the 
inferred column density (using {\tt TBABS} model in XSPEC12.4.0ai) is $N_H=1.6^{+0.6}_{-0.8}\times 10^{21}\cmdue$ 
($90\%$ confidence level), which is much smaller than the column density observed in outburst 
of $N_H=4.3^{+0.7}_{-0.5}\times 10^{21}\cmdue$ obtained fitting together Chandra HETG and RXTE/PCA 
data (Paizis et al. 2005). This value is also consistent with the Galactic value of 
$N_H=4.6\times 10^{21}\cmdue$. Here we adopt this value (see Table 2).

\begin{figure*}[htbp]
\begin{center}
\psfig{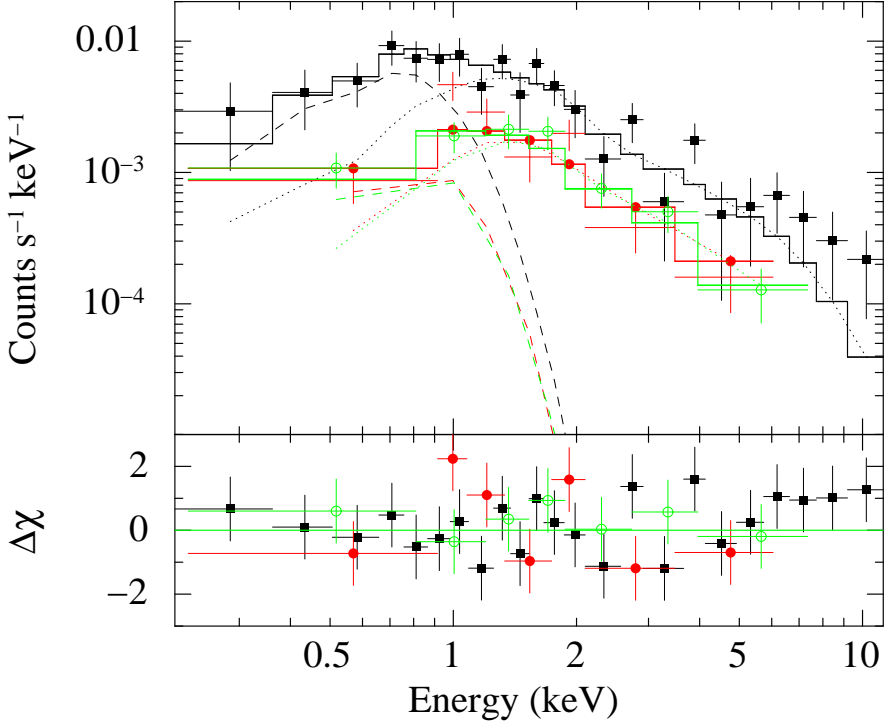}
\end{center}
\caption{XMM-Newton spectrum of IGR J0029. In the upper panel data
along with the best fit (absorbed) NSA plus power law spectrum are shown: pn data are 
marked with a open squares, MOS1 data with open circles and MOS2 data with filled
circles. In the bottom panel the $\chi^2$ contribution of each bin. Dashed and dotted lines 
refer to the neutron star atmosphere and power-law spectral components, respectively.}
\label{spe}
\end{figure*}

Fitting the data with the column density fixed at the above value provides relatively poor fits
(reduced $\chi^2=1.35$ in the case of a power law). Therefore we tried double component models.
Usually spectra of neutron star transients in quiescence are fit with a two components model: 
a neutron star atmosphere emission model (or a black body) plus a hard power law component.
These models provide a better description of the data. An F-test gives a $3.4\,\sigma$ probability 
for the significance in the fit improvement due to the extra model component (either black body 
or neutron star atmosphere, see Table 2).
This indicates that, if $N_H$ is held fixed at the value measured in outburst, the addition of a 
soft spectral component is significant. Estimates of the column density during quiescence 
of transient accreting millisecond X--ray pulsars are available only for SAX J1808.4--3658. For this source 
the quiescent value of the column density is consistent with the one observed in outburst (Heinke et al.
2007; Campana, Stella \& Kennea 2008).

\begin{table*}
\caption{Spectral fits.}
\begin{center}
\begin{tabular}{cccccc}
Model     & $N_H$             & Ph.Index          & Temperature          &$\chi_{\rm red}^2$& Flux$^b$ \\
          &($10^{21}\cmdue$)  &                   & (keV/log T)          & (dof) [nhp$^a$]  & (Luminosity)   \\
\hline
Power law &$1.6^{+0.6}_{-0.8}$&$1.7^{+0.2}_{-0.2}$& --                   & 0.89 (34) [66]   & 7.2 (7.7)\\
Black body&$<0.2$             & --                &$0.56^{+0.09}_{-0.07}$& 1.55 (34)  [2]   & 3.4 (3.6)\\
Power law & 4.6               &$2.5^{+0.3}_{-0.3}$& --                   & 1.35 (35)  [8]   & 8.5 (9.2)\\
Black body& 4.6               & --                &$0.38^{+0.09}_{-0.08}$& 2.86 (35)  [0]   & 4.4 (4.7)\\
NSA$^c$   & 4.6               & --                &$5.91^{+0.01}_{-0.01}$& 3.46 (35)  [0]   & 7.1 (7.7)\\
PL+BB$^d$ & 4.6               &$1.8^{+0.3}_{-0.3}$&$0.11^{+0.04}_{-0.03}$& 0.93 (33)  [59]  & 12.4 (13.3)\\
PL+NSA$^e$& 4.6               &$1.5^{+0.2}_{-0.5}$&$5.87^{+0.03}_{-0.04}$& 0.98 (34)  [49]  & 11.6 (12.4)\\
\hline
\end{tabular}
\end{center}

\noindent All quoted errors are at $90\%$ confidence level ($\Delta\chi=2.71$).

\noindent \tablenotetext{a}{Null hypothesis probability.}

\noindent \tablenotetext{b}{Unabsorbed 0.5--10 keV flux in units of $10^{-14}$ erg cm$^{-2}$ s$^{-1}$ and luminosity in units of $10^{31}$ erg s$^{-1}$ at 3 kpc.}

\noindent \tablenotetext{c}{NSA normalization fixed at $1.1\times 10^{-7}$ as for a source at 3 kpc. 
Neutron star mass, radius and magnetic field fixed to $1.4\msole$, 10 km and 0 G, respectively.}

\noindent \tablenotetext{d}{Black body radius $R=2.9^{+3.3}_{-1.7}$ km at a distance of 3 kpc. The black body
component comprises $36\%$ of the 0.5--10 keV unabsorbed flux.}

\noindent \tablenotetext{e}{The NSA component comprises $38\%$ of the 0.5--10 keV unabsorbed flux.}

\end{table*}

\section{Discussion and conclusions}

In recent years the quiescent properties of a handful neutron star transients 
have been studied in some detail thanks to dedicated observations mainly carried out with 
XMM-Newton and Chandra. `Classical' (i.e. not showing pulsations during outbursts) neutron star 
transients have quiescent (0.5--10 keV) luminosities within a narrow range of $10^{32}-10^{34}\ergs$ 
and their spectra display a soft component consistent with (cooling) emission from the entire 
neutron star and in some cases a hard (power-law) component contributing $10-50\%$ of the flux
(e.g. Campana et al. 1998). 
An increasing number of candidate neutron star systems has also been discovered in globular 
clusters (e.g. Pooley et al. 2003; Heinke et al. 2003). These are characterized by a soft spectrum 
consistent with emission from the entire neutron star surface, but do not display any evidence for 
a power law component (upper limits on the flux of such a component are very tight in some cases).
No outbursts have yet been detected from these sources. 
At the other extreme, there exist systems like PSR J0024--7204W in the globular cluster 47 Tuc, 
showing emission consisting of a dominant power law component, which is eclipsed for a portion 
of the orbit, and a soft thermal component, which appears to be persistent (Bogdanov, Grindlay 
\& van den Berg 2005). 

\begin{figure*}[htbp]
\begin{center}
\psfig{figure=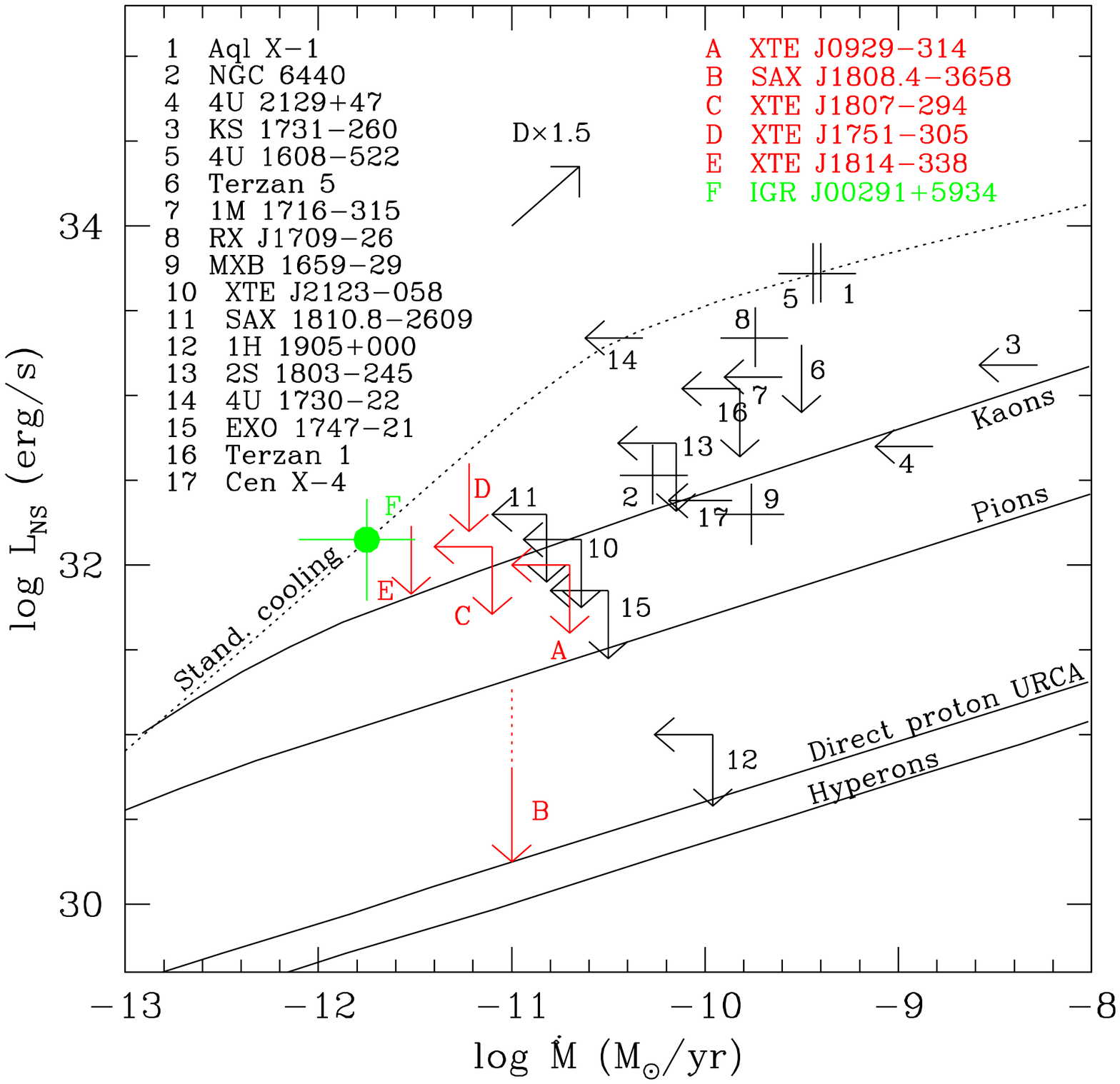,width=12truecm}
\end{center}
\caption{Cooling curves for various neutron star interior neutrino emission scenarios,
compared with measurements (or $95\%$ confidence upper limits) of the quiescent
bolometric luminosity and time-averaged mass transfer rate for several neutron star
transients (taken from Heinke et al. 2008). 
The big (green) dot indicate IGR J0029 values as estimated in this paper. Uncertainties have been 
accounted for by using a distance range of 2--4 kpc. 
The cooling curves are taken from Yakovlev \& Pethick (2004); 
the dotted curve represents a low-mass neutron star, while the lower curves represent high-mass 
NSs with kaon or pion condensates or direct Urca (Durca) processes mediated by 
nucleons or hyperons.}
\label{lc}
\end{figure*}

Transient accreting millisecond X--ray pulsars have low quiescent (0.5--10 keV) 
luminosities, below (and in some cases well below) $10^{32}\ergs$. In these systems the quiescent 
spectrum is made just by a power law component with no signs of a soft component. 
The best known system, SAX J1808.4--3658, shows a pure power law spectrum, with an upper limit
on the luminosity of the soft component of $\lsim 23\%$ (Heinke et al. 2007). 

Our XMM-Newton data on IGR J0029 in quiescence provides for the first time evidence for 
the presence of a soft component
in the quiescent spectrum of a transient accreting millisecond X--ray pulsar. The temperature
of this component ($k\,T\sim 70$ eV, when modelled with a neutron star atmosphere) is soft, but in 
line with other classical neutron star transients (e.g. Heinke et al. 2007). The bolometric neutron star 
atmosphere luminosity is $1.4\times 10^{32}\ergs$. This luminosity is usually ascribed to the heat generated 
deep in the crust during outburst episodes (i.e. accretion onto the neutron star surface) which is 
then radiated away through the atmosphere while in quiescence producing the quiescent X--ray emission 
(Brown et al. 1998; Haensel \& Zdunik 1990).
The efficiency of the re-radiation process depends on which fraction of heat escapes the neutron star 
as neutrinos rather than photons and in turn on the physical conditions (i.e., composition,
density, and pressure) of the neutron star interior. 

The accretion history of IGR J0029 is relatively well known. IGR J0029 was discovered during 
an outburst in Dec. 2004. The bolometric fluence of this outburst has been estimated in 
$1.8\times 10^{-3}$ erg cm$^{-2}$ using RXTE data (Galloway et al. 2005) or $1.37\times 10^{-3}$ 
erg cm$^{-2}$ using INTEGRAL data (Falanga et al. 2005). Thanks to the RXTE-ASM two
previous instances of activity, on 1998 November 26-–28 and 2001
September 11–-21 have been revealed (Remillard 2004). These results indicate that the
transient appears regularly, roughly every 3 yr. Indeed a new outburst started on Aug 13 2008 
(Chakrabarty et al. 2008), providing further evidence that the outbursts recur fairly regularly.
Assuming that all the outbursts have comparable fluence (as testified by similar outbursts peak in the
RXTE-ASM), one can estimate a long-term $F_{\rm acc}=1.4\times 10^{-11}\ergs\cmdue$.
Based on deep crustal heating theory, the quiescent flux is related to the long
term time-averaged flux $F_{\rm q}=F_{\rm acc}/135=1.0\times 10^{-13}\ergs\cmdue$ (Brown et al. 1998). 
This is in line with what observed. The goodness of this match lends support to the idea that 
the soft spectral component derives from cooling of the neutron star surface and not from accretion (since 
the spectrum at these low luminosities is expected to be soft, e.g. Zampieri et al. 1995).

Having estimated the quiescent cooling luminosity and the time-averaged flux (or mass inflow rate)
we can put constraints on models for the neutron star interior (Yakovlev \& Pethick 2004;
Levenfish \& Haensel 2007). As can be seen from Fig. 2, IGR J0029 lies on the standard cooling 
curve for a low mass neutron star. This is at variance with, e.g. SAX J1808.4--3658 and Cen X-4, 
for which low cooling luminosities have been inferred, hinting for enhanced neutrino emission produced
in the high-density core of a relatively high mass neutron star ($M>1.6-1.7\msole$, e.g. Colpi et al. 2001).

Our observation indicates that neutron stars showing coherent pulsation during their outbursts do not 
necessarily lack a soft component in their quiescent spectra. If the soft spectral component is due 
to deep crust heating emission and the presence or lack of this component is a proxy of its mass, as
massive neutron stars do not show it due to fast cooling. This would imply that light ($M\lsim 
1.6\msole$) neutron stars too can show coherent pulsations at the millisecond level, having accreted 
only $\lsim 0.2\msole$ (Lavagetto et al. 2004). 


\begin{acknowledgements}
SC thanks C. Heinke for providing the SM macro and data for producing Fig. 2.
\end{acknowledgements}

\end{document}